# Some aspects of resistive-to-normal state transition by direct and microwave currents in superconducting thin films with phase slip lines

O. G. Turutanov[1,2*], A. G. Sivakov[2], A. A. Leha[2], A. S. Pokhila[2], A. E. Kolinko[2], and M. Grajcar[1,3]

[1]Department of Experimental Physics, Comenius University, Mlynská dolina, 842 48 Bratislava, Slovakia

[2]B.Verkin Institute for Low Temperature Physics and Engineering of NAS of Ukraine, 47 Nauky ave., 61103 Kharkiv, Ukraine

[3]Institute of Physics, Slovak Academy of Sciences, Dúbravská cesta, Bratislava, Slovakia

*e-mail: oleh.turutanov@fmph.uniba.sk, turutanov@ilt.kharkov.ua

**Abstract**

Based on analysis of current-voltage characteristics and imaging of the resistive state of thin-film tin strips using the low-temperature laser scanning microscopy (LTLSM), the process of destruction of superconductivity by current and microwave irradiation with the formation and spatial rearrangement of the order parameter phase slip lines, and their transformation into discrete localized normal domains is shown. The prospects of LTLSM are considered form the point of view of study of the high-frequency properties of superconducting structures and spatial characteristics in the pre-critical state for instrumental applications.

## 1. Introduction

Superconducting electronics devices are typically manufactured using thin-film technology in an effort to reduce the size and increase the degree of integration of complex circuits, such as microwave Josephson parametric traveling wave amplifiers [1,2], digital SQUIDs [3,4], fast single-quantum logic circuits [5,6], registers and processors of superconducting supercomputers [7], voltage standards [8,9], spiral [10-12] and fractal resonators [13,14], superconducting nanowire single-photon detectors (SNSPDs) [15], etc. Knowledge of the distribution of superconducting currents and current density in various parts of the operating device is necessary for its design and obtaining its best parameters. It is especially important to know how close these dc and high-frequency currents, are to the critical value in order to avoid the appearance of local resistive states and unwanted losses, or even inoperability of the device. If the principle of operation of the device involves the transition of film elements into a resistive state, then the position of the place where the critical state is reached, and the transition from the superconducting state to the resistive one begins, is of great importance. Moreover, the occurrence of a critical condition may be completely undesirable, and then it is important to determine the "weak point". This spatial information cannot be obtained from measuring the current-voltage characteristics of the element under study. In addition, it is unrealistic to make electrical leads from each element to measure its current-voltage characteristics individually. In such cases, an adequate, spatially resolving research method is low-temperature laser scanning microscopy (LTLSM) [16].

In addition to applied interest, fundamental questions remain about the details of the process of destruction of superconductivity by current. At the dawn of superconductivity, there was a naive perception of transition from superconducting state directly to the normal one. Only much later sophisticated concepts of the resistivity appeared as a non-stationary and non-equilibrium process including magnetic flux creep, viscous and nonlinear flow of magnetic vortices, the emergence of phase slip centers and lines. Some processes in this context still elude complete understanding. Particularly, the attributes of localized states [17] that emerge during the transition from a superconducting channel to the resistive state under microwave irradiation remain unclear [18].

In this work, we will consider the emergence and development of the resistive state, mainly in the form of localized phase slip lines of the superconducting order parameter with further transition to the normal state, from the point of view of the applicability of the LTLSM method on both direct and high-frequency alternating currents. In the latter case, new opportunities open up to obtain maps of the spatial distribution of superconducting parameters in the subcritical state, not exceeding the critical current.

## 2. Phase-slip centers and lines in superconducting thin film strips with a uniform cross section

The resistive state of superconducting films is associated with two mechanisms, the magnetic flux vortices motion (dynamic mixed state) and the phase slips of the order parameter in localized domains, centers, lines and surfaces (PSC, PSL and PSS, correspondingly).

The SBT model [19,20] and later theoretical works [21-23], describe the structure of the resistive state in a quasi-one-dimensional superconductor with a transverse size of about coherence length $\xi(T)$ as localized nonequilibrium regions with a non-stationary order parameter at their center, oscillating with Josephson frequency. Excess quasiparticles appearing at the moments when the order parameter goes to zero at the center of the PSC, when its phase jumps by $2\pi$, diffuse in both directions, gradually coming into equilibrium with the superconducting condensate, where their electrochemical potentials are equalized. This means that there is a region around the PSC with the electric field inside within doubled electric field penetration length $2l_E$, determined by the diffusion velocity of quasiparticles and the relaxation time $\tau_Q$ of the imbalance of the electron-like and hole-like branches in the quasiparticle spectrum.

The formation of a next PSC while rising the transport current leads to a voltage jump, which is followed by linear section in the current-voltage characteristic (IVC) with discretely increasing slope, up to the normal resistance (Fig. 1).

As shown in [24,25], there is a resistive state in "wide" film strips with a width $w > \xi$ and even $w > \xi, \lambda$ (where $\lambda$ is magnetic field penetration depth in the thin film) due to phase slip lines (PSLs), which correspond to PSCs in quasi-one-dimensional channels. The theoretical description of PSLs is not sufficiently developed in comparison with PSCs, but experiments, including laser scanning microscopic ones [26] and observations of Shapiro steps [27], show that PSLs are also localized domains in which non-equilibrium non-stationary processes occur, similar to those in PSCs.

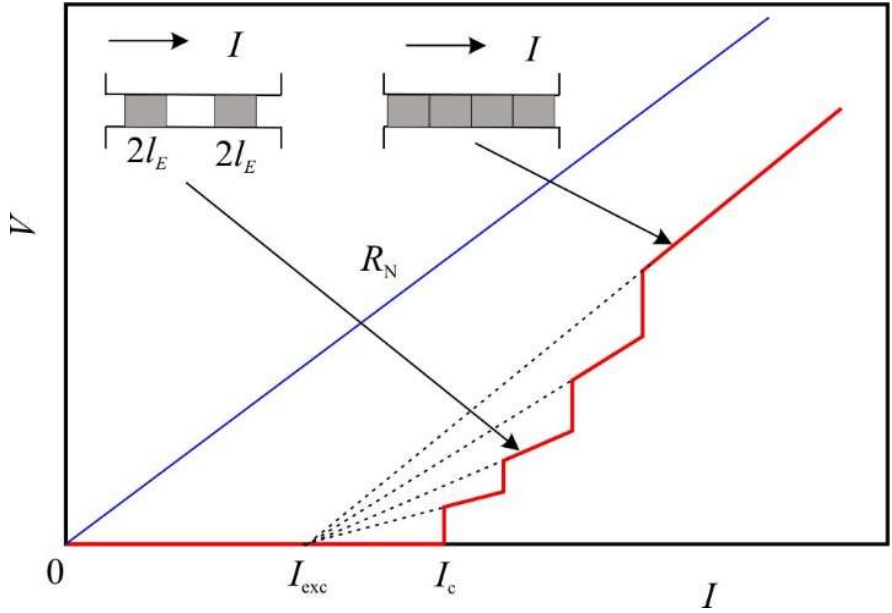

*Fig. 1*. Schematic IVC of a superconducting strip with current $I$, containing up to 4 PSCs or PSLs in its resistive state.

These structures are dynamic and essentially self-organized, current-induced "Josephson junctions". (Oscillations of the order parameter in them occur in a narrow spatial region of size $\xi(T)$). However, due to various depairing factors, such as temperature, parallel magnetic field, electric current, microwave field, injection of quasiparticles, etc., the order parameter oscillations can be suppressed, and resistive domains become normal. The formation of resistive domains and the nullifying the superconducting current are accompanied by instabilities and voltage jumps in the current-voltage characteristic (IVC). Even a thorough analysis of the IVC does not always give a complete view of the emergence, development and destruction of the resistive state without the use of spatially resolving methods like LTSLM.

Successive voltage jumps and linear sections whose differential resistance increases with almost the same increment, up to the normal resistance $R_N$ are observed in the IVCs of wide superconducting strips with uniform cross-section [19]. These linear sections are extrapolated to a point in the current axis at appr. $(0.5-0.6)I_c$. This value is called the excess current, and, according to SBT concepts [19,20], is the time-averaged supercurrent flowing through the PSC or PSL during the half-period of the order parameter oscillations. This structure is clearly visible in IVCs of relatively short strips (of several $l_E$) where only a small number of PSLs can fit. In long strips, the difference in the currents of formation subsequent steps becomes small and poorly distinguishable. Additionally, in the experiment it is not yet completely regular due to inhomogeneities of the material, cross-section and edges of the superconducting film, but the excess current is still observed up to currents several times greater than the critical one.

At high currents, the IVC becomes nonlinear, approaching $R_N$ due to the Joule heating over the bath (substrate) temperature with corresponding decrease in the excess current. However if measures are taken to actively stabilize the film temperature [28], the linear portion of the current-voltage characteristic can be significantly lengthened. Theoretically [23], "the current of instability of the normal state with respect to the formation of a superconducting nucleus," which is usually called the second critical current $I_{c2}$ and at which the IVC reaches the normal resistance line, can be ten times the critical current $I_c$. The same general picture with the formation of PSLs is valid for 2D variable cross-section thin-film structures (with current spreading), but with some distinctive features, as we will show below using the LTLSM. Moreover, when the superconductivity is destroyed by current in superconducting 3D structures such as point contacts, one should also expect the formation of phase slip surfaces [29].

This characteristic shape of the IVC well known to the researchers for many years, creates the illusion of easy proof of the occurrence of PSLs in certain specific cases. We will show below that this is not always the case, and for confident differentiation between PSLs, local normal domains and hot spots (the last two concepts are not the same), the LTLSM should be used.

The question of the formation of PSCs and PSLs by alternating microwave current [17] (with and without dc) with a frequency greater than the reciprocal charge relaxation time $\tau_Q$ still seems to be unclear [18,30]. The combined effect of the microwave field and direct current on a superconducting film strip is not additive. Below we discuss the experiments with the simultaneous action of direct and microwave currents on a superconducting strip.

In superconducting electronics circuits, the elements have a complex shape, which, taking into account the Meissner effect, skin effect and technological edge irregularities, makes it extremely difficult to calculate real current densities and identify "weak spots", especially at microwave frequencies, where multimode resonances and standing waves would arise. In these cases, it is possible to image, additionally to resistive regions, the current density in the subcritical state using vector non-contact measurement of the LTLSM response of superconducting elements at high frequencies [31].

## 3. Method of laser scanning microscopy of superconductors

The method of scanning the surface of a thin-film superconductor with a light spot at low temperatures to study the resistive state was proposed in [32] and belongs to the class of active probing techniques. The operating principle of LTLSM and a number of applications to superconductors are described in [17]. To understand the results presented in the proposed work, we briefly describe the method.

A laser beam is focused onto the surface of a thin-film superconducting sample under study using an optical system into a light spot of size limited by the optical limit. Typically a gas or semiconductor (diode) laser is used with radiation wavelength of about 0.63-0.67 μm, so the spot has a size of about 1 μm. The effect of light on a superconductor is mostly reduced to heating, although direct Cooper pair depairing is also observed, giving a small non-bolometric component [33]. The temperature rise in the spot center depends on the heat transfer from the film to the substrate. With the laser power of several milliwatts, it is 0.001 – 0.1 K. A polarizer installed in the optical path to obtain the spot illumination and thus local heating adequate for a particular experiment. For example, the transition width $R(T)$ in pure low-temperature superconductors can be 0.001-0.01 K, while in inhomogeneous HTSCs it reaches 0.1-2 K. The beam intensity is modulated by a sine or square wave, allowing lock-in response detection. The response can be any "global" property of a superconductor. In this work, we use "dc LTLSM", in which a direct current close to the critical one is passed through the sample, while the varying voltage that appears at its ends in the resistive state is the response.

Using mirrors with an electromagnetic drive installed in the optical path and controlled by software through a digital computer interface, the beam is deflected in mutually perpendicular directions, and the surface is scanned with a light probe forming a rectangular TV-like raster. By mapping the magnitude of the "global" response of the sample to the light probe coordinates, it is possible to construct an image of the LTLSM response, which is the basis for further analysis of the spatial properties of the superconductor. Since the response of a superconductor is nonlocal and determined by the structure of the resistive state, the map analysis is a complex task that requires a deep understanding of the characteristics of this resistive state and the geometry of the sample. Nevertheless, it enables one to obtain detailed information that is not available in other methods, for example, from the analysis of IVCs.

Returning to the question of the spatial resolution of the method, we should note that the size of the beam-heated spot with the local superconductor characteristics changed is larger than the light spot size. It is determined by the thermal healing length $\eta = (\kappa d / \alpha)^{1/2}$ (where $\kappa$ is the heat conductivity of the film, $d$ is the film thickness, $\alpha$ is the coefficient of heat transfer to the substrate) under constant illumination, but decreases with frequency when modulating the heat source [34], asymptotically approaching the size of the illuminated area. This makes it possible to increase the beam modulation frequency to the required value, taking into account the fact that the amplitude of the bolometric response also decreases with frequency. In some experiments, this serves as a tool for isolating the non-bolometric (non-thermal) response [34]. Typically, the beam modulation frequency is 10-100 kHz.

Response maps can be graphically represented in various forms, encoding the magnitude of the response in shades of gray, conditional colors, and height in a "mountain landscape" representation. In this work, we use a grayscale representation, where the minimum response level is black and the maximum response is white. To analyze the evolution of the resistive state under various conditions, it is sufficient to know the relative (normalized) response values, so we do not present the absolute response values here.

Note that the response can be any measurable quantity, depending on the superconducting characteristics of the sample, and the current can be alternating, including high frequency. In the latter case, it becomes possible to make non-contact measurements and separate the active and reactive components of the response using lock-in technique, e.g. to test elements of superconducting microwave electronics directly in their operating mode [17, 33].

## 4. Experimental technique

The samples under study are thin (30-200 nm) tin films deposited on a single-crystal quartz substrate by thermal evaporation in a vacuum of $10^{-6} - 10^{-7}$ Torr. For electrical and LTLSM four-probe measurements, the samples were patterned into short and long strips with wide current pads for current input and narrow potential leads. (Current spreading effects with two-probe connections are also discussed in Section 5.) The samples were patterned using electronic lithography or cutting with a focused ultraviolet laser beam. The IVCs were recorded forward and backward in two polarities, but where this does not matter, one branch is shown in the figures.

LTLSM experiments were performed in an optical helium cryostat in the Earth's magnetic field. The scanning area slightly exceeded the dimensions of the sample between the potential contacts. However, the recording time for one full frame was 5-15 minutes due to the averaging of the response signal at each measurement point, so the recording time for a long image series could reach several hours, which exceeded the operating cycle time of the cryostat (about 8 hours). In these cases, a narrow area in the center of the sample was scanned, taking into account the symmetry of the state, namely, the fact that the PSLs intersect the entire cross section of the sample perpendicular to the line of the current flow.

The wavelength of the laser diode radiation was 0.63 µm, the initial power was 2 mW, which was modulated at frequency of 10-100 kHz by the supply current and attenuated by a polarizer. The power incident on the surface of the sample was not measured, but in each specific experiment it was selected based on the minimum impact that would give a response above the background noise. Typically, the change in voltage at the ends of the sample was 1-5% of the dc voltage across the unilluminated sample. The ac voltage component was amplified by a lock-in amplifier. The absolute values of the ac voltage response (typically 2 µV - 200 µV are stored in data files, but it is not practical to display them on the response maps, because the goal of LTLSM experiments is, as a rule, qualitative visualization of changes in the spatial structure of the resistive state of a superconductor. To obtain numerical local values of the superconducting parameters of the sample, it is not sufficient to measure carefully all the irradiation and heating parameters (incident power, spot size, reflectivity, heat transfer coefficient to the substrate and thermal conductivity of the film). A theoretical model of the resistive state of the superconducting film is needed, which would give the spatial distribution of superconducting and normal currents provided the Abrikosov vortices, phase slip lines and normal areas may be simultaneously present in a specific situation. Such a complete theory is a major challenge and is currently lacking. This creates certain difficulties in interpreting the results, which is based on extensive experience with LTLSM.

## 5. Experimental results and discussion

The main material of modern lithographed superconducting microstructures and integrated circuits is niobium, but for the purposes of studying critical current and resistive states, so-called "soft" low-temperature superconductors, mainly indium and tin, have proven themselves well. They are easily obtained in the form of thin films of stable quality in a relatively low vacuum, and have fairly large (several microns or fractions of microns) coherence lengths and penetration depths of magnetic and electric fields near the critical temperature. The inelastic relaxation time $\tau_\varepsilon = 10^{-9} - 10^{-10}$ s, which ultimately determines nonequilibrium processes in the superconducting state, lies in the well-developed microwave frequency range, while the charge imbalance relaxation time $\tau_Q$ corresponds to even lower frequencies. These "classical" superconductors are much more convenient for studying the fundamental properties of superconductivity, especially using spatially resolving methods, and for modeling the behavior of other superconducting materials, including niobium and HTSC, whose characteristic superconducting lengths are much less than a micron.

A number of instruments such as bolometers, film magnetometers, etc. use the resistive transition curve $R(T)$. At the same time, when measuring $R$, the majority of experimenters actually record the voltage drop $V$ at a given transport current $I$ and consider $R = V/I$. But when the sample transits to the superconducting state, the transport current is divided into normal and superconducting components, and the voltage drop can also have a different nature due to the longitudinal electric field in the normal regions in which quasiparticle diffusion occurs and the change in the phase of the order parameter over time.

Fig. 2 shows set IVCs of a thin-film superconducting tin strip of $30 \times 100$ µm in size, taken at different temperatures during the superconducting transition. The 200 nm-thick film is deposited by thermal evaporation in a vacuum onto a single-crystalline quartz substrate, patterned by electron lithography, the resistance ratio R$_{300}$/R$_{4.2}$=20.

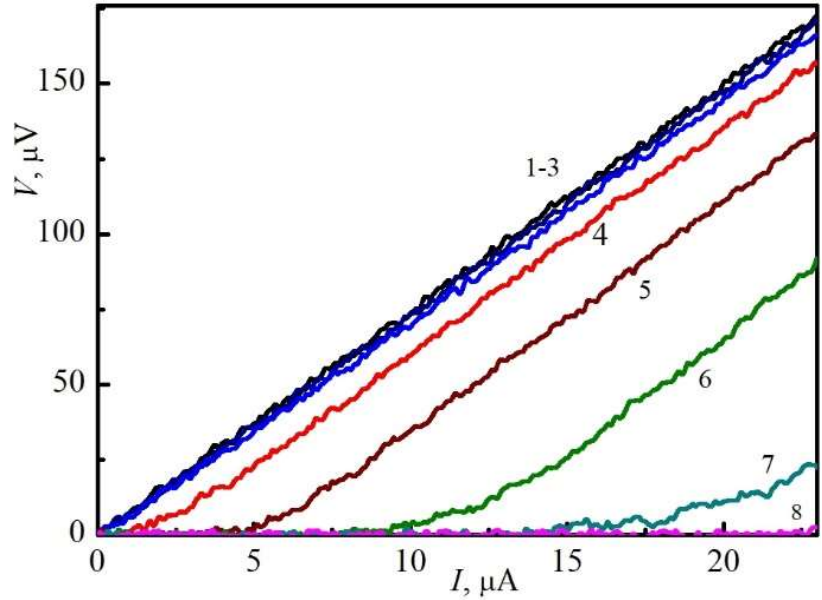

*Fig. 2.* Set of IVCs of a tin strip at different temperatures. Strip size is $30 \times 100$ μm Temperature for curves: 1 – 3.757 K, 8 – 3.738 K, temperature decrement 0.003 K.

The initial sections of the IVC were recorded with a temperature step of 0.003 K. One can see that the transition to the superconducting state, by criterion of the appearance of a noticeable critical current (above the measurement noise threshold), occurs in a narrow temperature range of about 0.003 K (from curve 3 to curve 4), and then I-V curves show the critical current. Note that the I-V curves run parallel to the line of normal resistance $R_N$ with excess current.

The excess current $I_{exc}$ indicates the presence of oscillations of the order parameter phase over time in the sample. The modulus of the order parameter becomes zero at certain moments of the cycle, then the phase changes (slips) by $2\pi$, while the voltage drop in the nonequilibrium region around determines the frequency of these oscillations according to the Josephson relation. At this time, the normal component of the current flows. During the rest of the oscillation period, a supercurrent flows, the averaging of which over time gives the excess current. Thus, a non-zero excess current witnesses that alternating currents of Josephson frequency flow in the PSL region. This scenario is confirmed by the presence of Shapiro steps in the IVCs of the bridge when it is under microwave irradiation. Fig. 3 demonstrates the behavior of Shapiro steps in the IVCs of $10 \times 25$ μm tin bridge at various microwave irradiation powers. The position of the voltage steps corresponds to the microwave radiation frequency (1 GHz) and its harmonics. The critical and excess currents fall with increasing irradiation power, and the steps disappear since the moment $I_{exc} = 0$. This means that PSL turns into a localized normal domain (we will abbreviate it as NLD for better pronunciation), in which oscillations of the order parameter no longer occur, while the NLD resistance remains approximately equal to the PSL resistance. Note that this normal domain is non-equilibrium and exists only in current-driven state.

Decrease in temperature and increase in the critical current causes a relatively large (about 0.01 K [28]) local overheating and hysteresis when a PSL emerges. Since the excess current depends on temperature, it is possible to find a temperature at which the occurrence of a PSL leads to local heating in its core that makes the excess current zero. Microwaves works in the same way, causing the formation of a NLD (see Fig. 3) with zero excess current, which in [18] is interpreted as ac PSC.

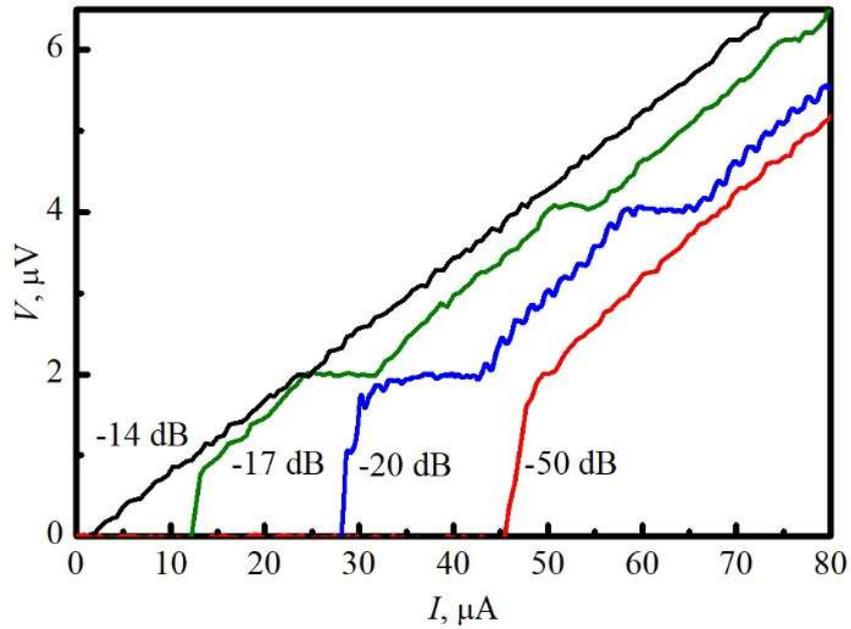

*Fig. 3*. IVCs of a tin bridge with PSL and Shapiro steps under microwave irradiation. Microwave frequency 1 GHz. Bridge size is $10 \times 25$ μm. Relative power levels are indicated in the curves.

Fig. 4 shows a series of IVCs demonstrating the transition of the PSL to the NLD with rising the critical current. In this experiment, the critical current increased due to the effect of stimulation of superconductivity by microwave radiation.

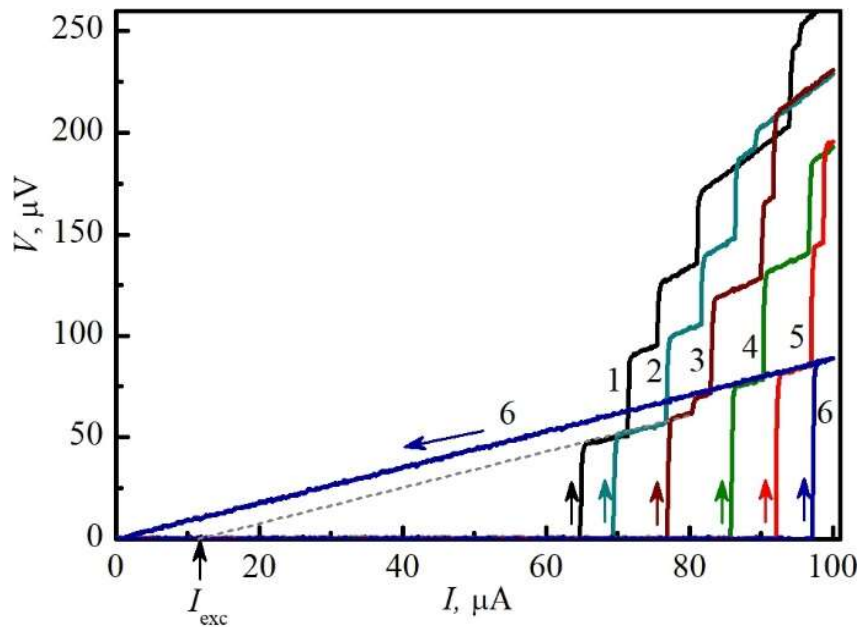

*Fig. 4*. IVCs of 0.5x50 μm tin strip at different levels of microwave irradiation, illustrating the transition from dc PSL to NLD. Temperature 3.9 K, irradiation frequency 19.2 GHz, incident powers for curves 1-6, respectively, -10.3, -9.8, -9.0, -8.2, -7.6, -7.1 dBm. The arrows indicate the curves recording direction. $I_{exc}$ denotes the excess current for the first three curves.

Curve 2 indicates that both temperature and microwave have a depairing effect. Up to a power of -9.0 dBm (curves 1-2), there is a jump in IVC from the critical current to the first step with excess current, but with increasing current at the same power it turns into an NLD with $I_{exc} = 0$ (curve 3). With further increase in power, the domain remains in the normal state with almost the same resistance (curves 4-6). The backward branch of the IVC goes to the coordinates origin (Ohm's law). Finally, the critical current drops to zero at the critical power, and the IVC becomes ohmic (NLD) (see Fig. 3, curve -14 dB).

After the superconducting condensate reaches the depairing velocity, the resistive state develops in a wide currents range, which is accompanied by an increase PSLs in the number, tending to uniformly fill the entire volume of the sample [27] provided that it is highly homogeneous. In real samples, the local values of the critical current have a scatter due to the geometry of the sample, any kind of inhomogeneity of the material, the order parameter and the edge barrier. At low temperatures, PSLs can be "pinned" by thermal inhomogeneities. In these cases, the addition of new PSL occurs alternately on both sides of the already existing resistive region, since the order parameter near the boundary is smaller than in the rest of the homogeneous sample due to heating. This process of spatial development of the resistive domain with the addition of anPSL is shown in Fig. 5.

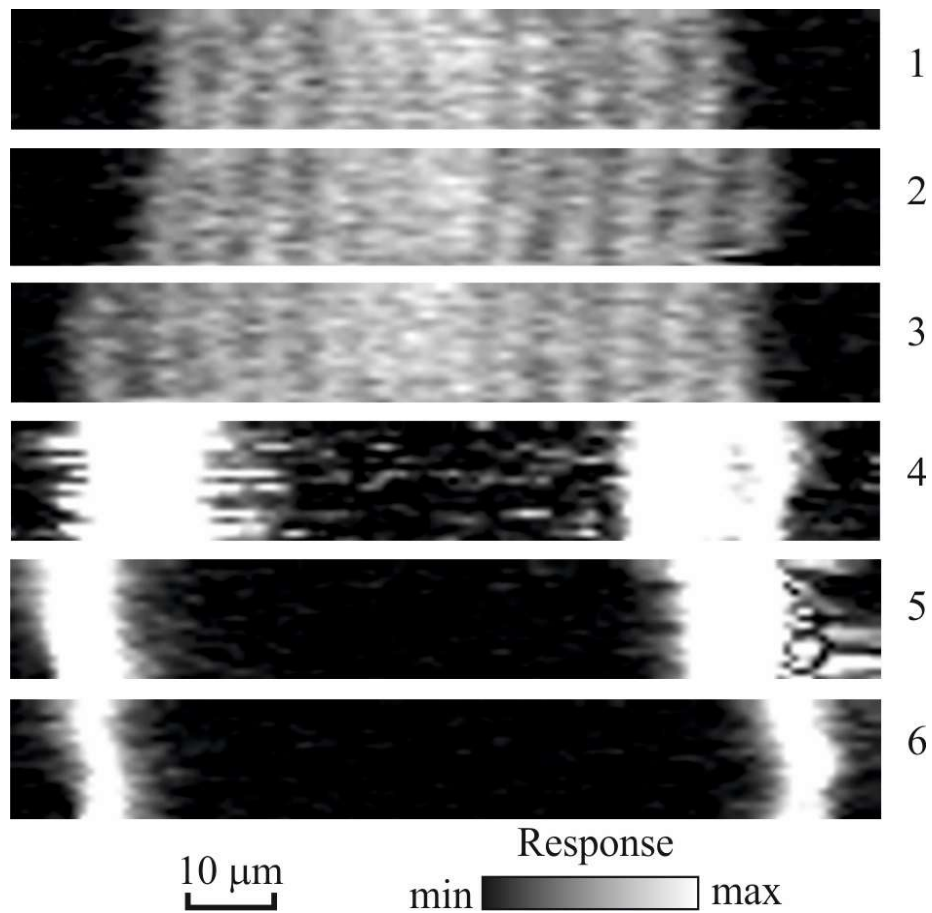

*Fig. 5.* An increase PSLs in the number in a thin film strip with increasing current and the propagation of the normal domain. Temperature is 3.64 K, current values for successive scans are:. 1 — 0.53 mA, 2 — 0.56 mA, 3 — 0.58 mA, 4 — 0.63 mA, 5 — 0.65 mA, 6 — 0.68 mA. Beam frequency modulation is 17.5 kHz. Grey-scale maps show LTLSM ac voltage response. Brightest regions correspond to maximum voltage ( 2 μV ) while darkest to the minimum ( 0 μV ).

PSLs look like light stripes in the cross section of the sample. The response from the PSL originated due to modulation of the excess current in the PSL core by heating with the light spot and the direct depairing effect of the photons. When the entire sample area is filled with the PSLs, the temperature heterogeneities are smoothed out, since the thermal healing length in the film, as a rule, is greater than the characteristic PSL size ($\eta > 2l_E$). With increase in the dissipated power, the sample temperature rises and the excess current drops down. When the order parameter in the PSL core becomes zero, the PSL transforms into the NLD. Superconducting and normal currents flowed alternately in the PSL core until this moment, but then the transport current becomes normal in the NLD. The excess current $I_{exc}$ vanishes in individual PSLs while NLDs appear instead, and, as a result, the normal region consists of a discrete set of individual NLDs (the last three scans in Fig. 5). This transition begins at the center of the sample, since its central part has higher temperature than at the edges due to improved heat sink into wide metal current leads. The details of the transition in the middle of the sample are lost because of large response range when passing from scan 3 to scan 4. Note that the wide white lines in the last scans correspond to several PSLs adjacent to the S-N boundaries at the edges of the sample. There are 6 of them in total on scan 4, 5 PSLs on scan 5 and 2 PSLs on scan 6.

NLDs are not "hot spots" in the classical sense: with the set current (i.e., with high internal resistance of the current source), the hot-spot region, overheated above $T_c$, should spread over the entire sample. In our case, this domain is spatially stable with the set current, and the size of the normal region increases discretely as the PSLs transform into NLDs. It is difficult to counter the transitions of individual PSLs into NLDs one-by-one given the small scatter of local superconducting parameters. However, if one artificially creates one strong inhomogeneity, or an S-N boundary, then the process of formation of the PSLs can be "stretched" in current, as in [35]. The suppression of the order parameter can be made by, for example, microwave power rather than self-heating by the transport current.

To image the process of transition of a sample to the normal state via formation of NLDs, we used the sample configuration shown in Fig. 6(b). We created an inhomogeneity near one of the potential contacts in the form of a narrow cut, thereby locally lowering the critical current and expanding the current range of the transition. To obtain spatial response maps, we scanned the central part of the sample in the form of a rectangular raster with dimensions slightly larger than the distance between potential contacts and a width less than that of the sample. This was done in order to avoid creating thermal inhomogeneities with the beam at the edges of the sample, which would dramatically change critical conditions for the whole sample. In addition, reducing the raster size significantly shortens scanning time to obtain a large series of measurements (47 scans), taking in mind that the spatial characteristics of resistive structures are uniform along the cross section perpendicular to the current.

It was shown earlier [18,30] that in the presence of a stepwise structure in the IVC, an increase in the microwave irradiation power leads to a decrease in the currents of formation of individual steps. In this case, the values of the critical power sufficient to nullify the critical current and the currents of separate steps with the formation of NLD are different and increase with number of the next step.

Fig. 6(a) shows the IVC of a film strip $28 \times 70$ μm in size irradiated by microwaves with critical power at which the first linear ohmic section appears in the IVC, i.e. the critical current vanishes. Meanwhile, the current of formation of subsequent steps remains finite.

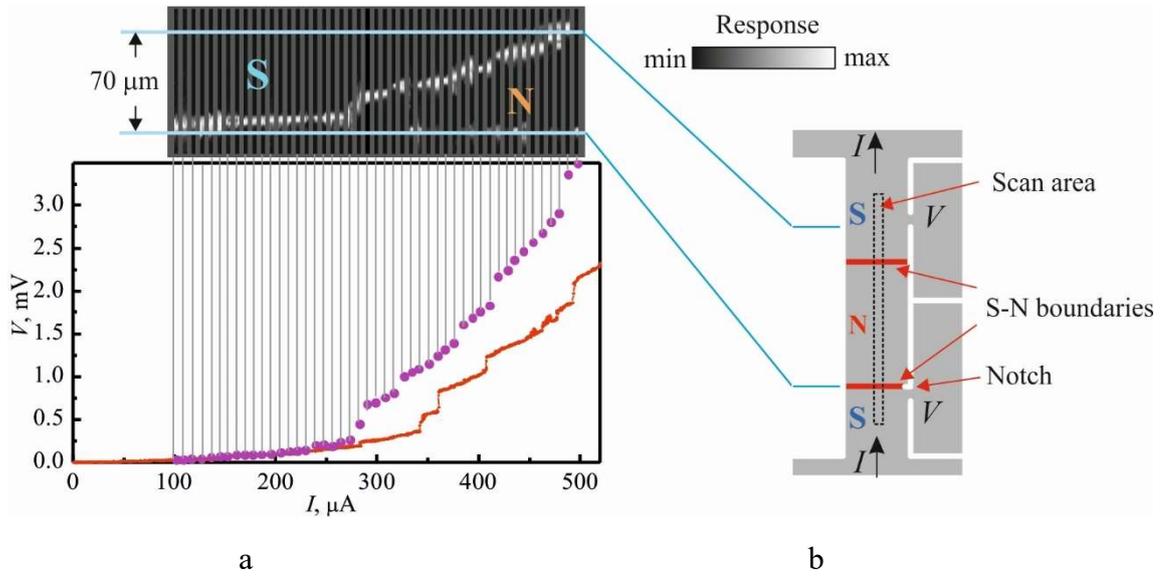

*Fig. 6.* Propagation of a normal domain with increasing direct current in a superconducting tin strip 28x70 μm in size, irradiated with microwave radiation of "critical" power (a). The solid red curve shows the initial IVC without laser illumination. The purple points on the IVC during laser probing indicate the currents at which 47 LTLSM scans along the sample were obtained. Blue lines indicate the position of potential leads in the scanned sample. The contrast of LTLSM images is enhanced and noise is removed. Film thickness is 30 nm, temperature is 3.85 K, laser beam modulation frequency is 18.9 kHz. The diagram of the sample (b) shows cuts (white) in the film (gray), creating a strip configuration with wide banks for the current supply ($I$) and potential leads ($V$), from which the response signal was taken. The dotted rectangle inside the strip indicates the scanning area.

When scanning the sample surface with a laser probe at different currents (Fig. 6(a)), S-N boundaries are imaged on the response maps. Individual scans correspond to the points in the IVC during scanning, which differ slightly from the pre-recorded IVC due to heating by the laser probe. The signal appears only at the interface between the N and S regions due to modulation by the laser probe of the magnitude of the electric field jump changing due to the effect of Andreev reflection from the S-N interface on a scale of the order of the coherence length $\xi(T)$ [35 and references therein]. The S-N boundary, "pinned" by the cut near the potential lead (lower lead in Fig. 6(b)), goes beyond the potential lead, slightly contributes to the response voltage, and is therefore almost invisible on the scans.

As the current increases, steps appear in the IVC, which indicate the shift of the S-N boundary on the rasters at corresponding current values. This means that the stepwise structure of the IVC is associated with discrete advancing of the normal region. This does not follow from the IVC shape, which at first glance can be (erroneously) considered as an evidence of the PSL formation. This is the danger of hasty interpretation of experimental data based only on the shape of an IVC.

A smooth increase in microwave power leads to an increase in the number of linear discrete ohmic sections in the IVC, corresponding to the sequential PSL-to-NLD transformations (Fig. 7(a)). The discrete set of linear ohmic sections in the IVC at certain microwave powers corresponding to the critical power of individual steps shows that the position of the S-N boundary changes in jump-like manner with the addition of the next NLD, i.e. the propagation of the normal region is discrete. This evolution occurs within very narrow power range of 2 dB (just 1.6 times). Sharp dropdown of the critical current at a certain power limiting its growth due to stimulation of superconductivity by microwave field, has long been noted by experimenters [36].

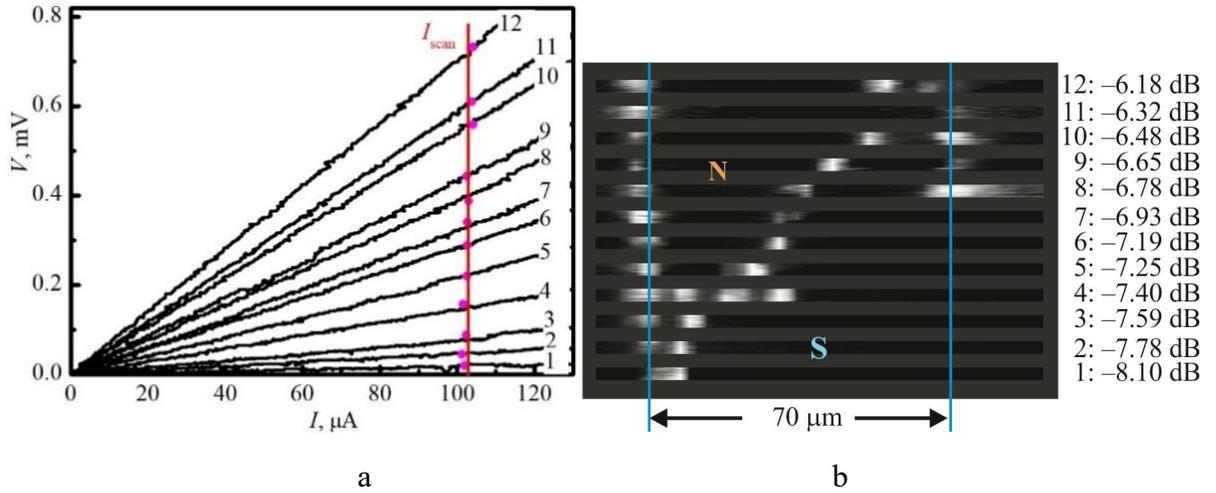

*Fig. 7*. IVC (a) during the propagation of the S-N boundary in superconducting tin strip $28 \times 70$ μm in size under the microwave irradiation of various power. Relative power values are indicated next to the scans (b). The red line in panel (a) indicates the current at which the sample was scanned, the purple dots are the actual measured current values. The blue lines in panel (b) indicate the position of potential leads in the scanned sample. The contrast of LTLSM images is enhanced and noise is removed. Light areas correspond to larger response amplitudes. Temperature is 3.85 K, laser beam modulation frequency is 18.9 kHz.

Above, we considered the development of resistive state with increasing current up to the transition to the normal state in film strips with constant cross-section along the length, and hence the current density. In this case, the PSLs have the form of straight lines intersecting the strip at right angles. The question arises of how resistivity develops in a superconducting structure with a nonuniform current density.

Fig. 8(a) presents the shape of the thin film tin microbridge (Dayem bridge), 10 μm in width, patterned by UV laser cutting. In the normal state, current flows into the bridge from wide banks, as shown in Fig. 8(b), and the potential leads measure the voltage only over the bridge, excluding the banks. Having cooled the bridge below the superconducting transition temperature, we drive it to the normal state with current (Fig. 8(c). The current density quickly falls down in wide banks due to spreading, therefore, the current becomes superconducting at distance $l_E$ from the bridge (more precisely, from the S-N boundary it created), and conditions emerge for the formation of pair of phase slip lines on both sides of the central cut. PSLs are shaped as the lines of equal current density (practically circular arcs) having length of $2l_E$ along the current flow direction. As soon as, with a further increase in the current, at a distance 2lE from the first line, the critical current density is reached, the next LPF comes to existence, and so on.

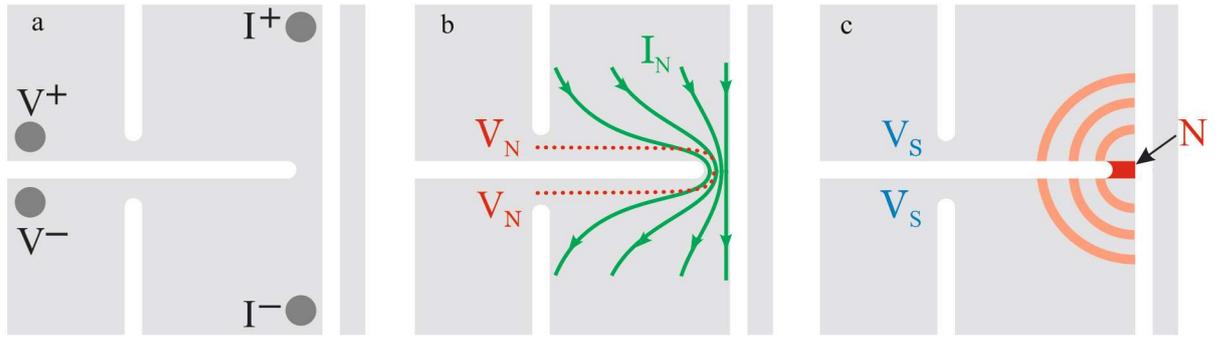

*Fig. 8.* (a) Four-probe measurement bridge configuration; (b) Sketch of current spreading in the normal state of the film. Potential leads measure the voltage at the constriction itself, excluding the banks; (c) Scheme of the PSLs formation in the superconducting banks, with the bridge driven to the normal state by current.

The appearance of three PSLs one-by-one in the banks of the bridge with current rise is imaged on a series of the LTLSM maps (Fig. 9).

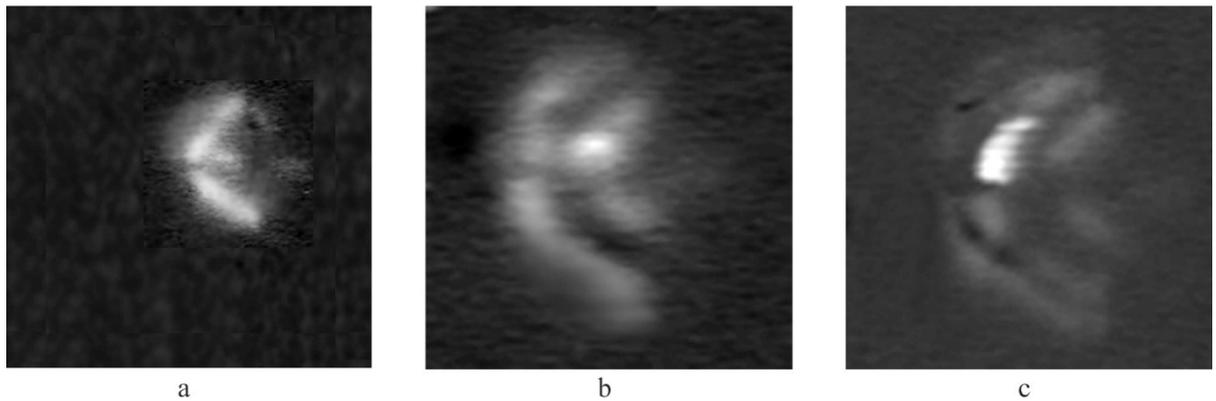

*Fig. 9.* PSLs in the wide banks of thin-film microbridge with increasing current: (a) 0.51 mA, (b) 0.58 mA, (c) 0.61 mA. The bridge width is 10 μm. Raster size is 60x60 microns. Temperature is 3.78 K, beam modulation frequency is 2.2 kHz.

All subsequent PSLs have larger width than the previous ones, with the same length $2l_E$ in the current flow direction, and therefore make a smaller and smaller contribution to the differential resistance. (The length of the LPF means the same as in narrow channels - the length of the localized nonequilibrium region along the current flow; the width, respectively, is the size in the perpendicular direction, although this somewhat contradicts the visual image).

Superconducting potential leads measure the voltage across the bridge in its normal state, plus the voltage drop across each of the PSLs formed in the banks as the current increases. Then the expected IVC should have the form of initial ohmic part and then the usual voltage steps and inclined linear sections with excess current. The difference from the IVCs of strips with uniform longitudinal current distribution is that the increments of the differential resistances of the linear sections are not the same, but decrease for higher step (PSLs) number. We can observe exactly this IVC in Fig. 9 from recent work [37] for bulk NbSe$_2$-Cu point contact. It is explained by the sequential formation of phase slip surfaces (PSS, [29]) in the superconducting bank around the point contact of the superconductor with the normal metal (copper). As we can see, the resistive state of a bulk point contact can be modeled and imaged with a thin film bridge having wide banks.

## 5. Conclusion

The mechanism of resistivity driven by the process of slippage of the order parameter phase in current-loaded superconductors is universal. This mechanism follows the dynamic mixed state (motion of Abrikosov vortices) where applicable and spans a wide range of currents up to the ultimate transition to the normal state.

When measuring the resistance of superconductors in the resistive state, even at low currents, it is necessary to consider that non-equilibrium non-stationary processes occur near the superconducting transition. They lead to the coexistence of normal and superconducting components of the transport current, which contribute differently to the measured voltage. A more adequate method is the measurement and analysis of I-V characteristics. However, I-V characteristics do not always provide the correct understanding of the resistive state's nature in each specific case. Spatially-resolved low-temperature laser scanning microscopy helps to determine the structure of the resistive state in superconducting films and to explain the peculiarities of the I-V characteristics, which could otherwise be misinterpreted. In particular, discrete normal domains (NLDs) arise and multiply in strong electromagnetic microwave fields that destroy superconductivity, being originated by and localized at their size, which is not obvious from the IVC shape.

It has long been understood that phase slippage is inherent not only to quasi-one-dimensional superconducting channels but also to 2D and 3D superconducting structures. The spatial patterns of the resistive state will differ in structures having longitudinally inuniform cross-section, current density or critical current density. In many cases, they can be imaged using low-temperature laser scanning microscopy. The I-V characteristics are also different for them. For example, in Dayem bridges, point contacts, etc. with strong current spreading the voltage steps in the IVCs corresponding to the formation of the next phase slippage area (PSL, PSS) become highly non-equidistant in terms of current, and the differential resistance also increases non-linearly with the step number. This work shows an example of imaging the resistive transition by current in such a structure.

We demonstrated how LTLSM technique helps in determining the structure of the resistive state in thin films when superconductivity is being destroyed by current, in addition the IVC analysis.

Finally, we would like to note that the progressing method of microwave LTLSM, which can be entirely non-contact, enables monitoring spatial changes in local impedances and currents in superconducting structures like high-temperature superconductor (HTS) resonators and SQUID-based metamaterials [16,31]. With vector measurements (lock-in technique) in the gigahertz range, the method is capable of separating the active and reactive components of the response. The inductive component reflects the magnitude of the superconducting current, that can be used to identify the weak spots in subcritical states without transiting to the resistive domain. In addition to the obvious technical interest of testing superconducting microwave devices at their operating frequencies, this method can provide experimental answers to fundamental questions about the structure of the resistive state at frequencies corresponding to characteristic relaxation times in superconductors.

**Acknowledgments**

The work was supported in part by grant of the National Academy of Sciences of Ukraine UA0122U001503, SPS Programme NATO grant number G5796, grant of Slovak Republic government (application number 09I03-03-V01-00031).